# Parametric wave mixing enhanced by velocity insensitive two-photon excitation in Rb vapour

ALEXANDER M. AKULSHIN,[1, *] DMITRY BUDKER,[2,3] RUSSELL J. MCLEAN[1]

[1]Centre for Quantum and Optical Science, Swinburne University of Technology,
PO Box 218, Melbourne 3122, Australia
[2]Department of Physics, University of California, Berkeley, CA 94720-7300, USA 3 Johannes Gutenberg
[3]Johannes Gutenberg University, Helmholtz Institute, D-55128 Mainz, Germany
*Corresponding author: aakoulchine@swin.edu.au

We demonstrate how the orientation of the Rb cell can significantly affect the intensity and spectral characteristics of both the frequency up- and down-converted fields generated by nonlinear processes in Rb vapour. The efficiency of parametric wave mixing in Rb vapour excited to the $5D_{5/2}$ level by two-colour resonant laser light can be significantly increased by seeding the excitation region with coherent and directional radiation at 5.23 μm that is resonant with the population inverted $6P_{3/2}$ -$5D_{5/2}$ transition. It has been shown that the process of velocity insensitive two-photon excitation is central to understanding the observed coherent blue and mid-IR light enhancements and that the velocity insensitive and velocity selective two-photon excitations could produce two co-existing but spectrally distinguishable mid-IR fields at 5.23 μm.

http://dx.doi.org/10.1364/AO.99.099999

## 1. INTRODUCTION

Nonlinear processes in atomic media can generate optical fields with substantially different wavelengths to those used to excite the atoms. Parametric four-wave mixing (FWM) of low power cw resonant laser fields in Rb and Cs vapours has been shown to be responsible for the generation of directional blue and IR light [1-11]. The new optical fields may be highly coherent [12, 13], making this method of generation of polychromatic optical fields useful for high-resolution applications. The technique is also potentially important for low atom number detection [14] and for quantum information science [15], both of which exploit the conversion of light from one spectral region to another.

Studying the coherent blue light (CBL) generation in Rb vapour excited by co-propagating laser light at 780 and 776 nm, as Figure 1 illustrates, we have found that the efficiency of FWM depends, among other factors, on the geometry of the optical arrangement, including the orientation of the vapour cell relative to the applied laser beams. CBL and mid-IR radiation are usually stronger when the cell windows are nearly perpendicular to the applied laser beam. In addition, the observed intensity enhancement of both frequency up- and down-converted radiation is accompanied by irregular distortions of their spectral dependences. Figure 2 shows that the smooth spectral profiles of forward-directed mid-IR radiation at 5.23 μm and CBL obtained at relatively large cell tilt (θ ≈ 70 mrad) become noisy and temporally unstable when the cell window is nearly normal to the applied laser beam (θ ≈ 5 mrad, where θ is the angle between the incident beam and the normal to the window). It was also noticed that the spectral profile of the generated 420 nm light depends on the position of the laser light focusing spot inside of the cell [11].

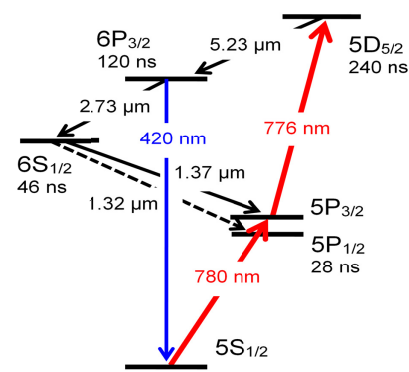

Fig. 1. Rb atom energy levels involved in the $5S_{1/2}$ - $5D_{5/2}$ two-photon excitation and subsequent parametric FWM processes.

It is apparent that these distortions are due to reflections from the cell windows that direct light back through the cell. However, at the resonant condition for the parametric FWM process the reflected beam may contain optical fields at several wavelengths. In addition to the attenuated laser light at 780 and 776 nm it may contain collimated blue light and a number of IR fields at 1.32 μm, 1.37 μm, 2.73 μm and

5.23 μm [1-4]. We emphasize that the FWM process that generates the CBL involves co-propagating light at 780 and 776 nm and 5.23 μm, required by the phase matching condition. Reflected light cannot directly participate in the process of parametric FWM in the forward direction. Nevertheless, its presence is causing a significant enhancement of the generated fields in that direction. Here we address which of these reflected fields is responsible for the marked impact on the nonlinear wave mixing and the mechanism for the observed sensitivity to Rb cell orientation.

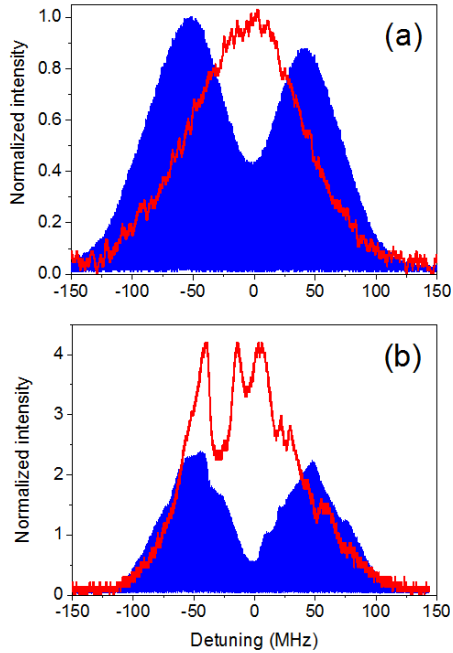

Fig. 2. (a, b) Intensity of forward-directed blue (blue profiles) and mid-IR radiation at 5.23 μm (red profile) at two different cell tilts, 70 mrad and 5 mrad, respectively, as a function of the 776 nm laser detuning from the 5P$_{3/2}$(F'=4)-5D$_{5/2}$(F''=5) transition, while the fixed frequency 780 nm laser is tuned to the 5S$_{1/2}$(F=3)-5P$_{3/2}$(F'=4) transition. The intensity of the coherent blue and mid-IR radiation at the small cell tilt is normalized to the corresponding intensities obtained at larger cell tilting.

First, we conduct an experimental test that is based on the idea of seeding radiation, assuming that the details of the mixing process can be inferred by analysing the CBL intensity changes caused by injecting extra resonant light at different wavelengths. Next, we describe the appearance of two spatially and spectrally distinct forward-directed fields at 5.23 μm originating from velocity selective and velocity insensitive two-photon excitation [17].

## 2. EXPERIMENTAL SETUP

The experimental arrangement for frequency up- and down-conversion using low power cw laser radiation is essentially the same as that described in our previous publications [4, 16, 17]. A 5-cm long glass cell with 0.96 mm-thick sapphire windows is used in the experiments. The transmission of a 1 mm-thick sapphire plate is approximately 60% at 5.2 μm [18]. A specially designed resistive heater for the cell provides the temperature gradient required to prevent Rb atoms from condensing on the cell windows. The Rb atom number density is estimated based on experimentally recorded linear absorption profiles for the Rb D1 line and specialized software [19].

The optical scheme of the seeding experiment is shown in Figure 3. The Rb cell tilt (θ ~ 70 mrad) is large enough to avoid the above-mentioned distortions. The polychromatic radiation is reflected toward the Rb cell by a gold-coated glass plate in such way that the incident and reflected beams are spatially separated inside the cell by a few mm.

The spectrum of the reflected radiation inside the cell can be controlled by inserting different optical filters. A blue glass filter transmits light at 420 nm with approximately 10% attenuation and completely blocks both the laser and IR radiation. A thin uncoated glass plate is transparent for the blue and laser light (less than 10% attenuation) but totally absorbs radiation at 5.23 μm.

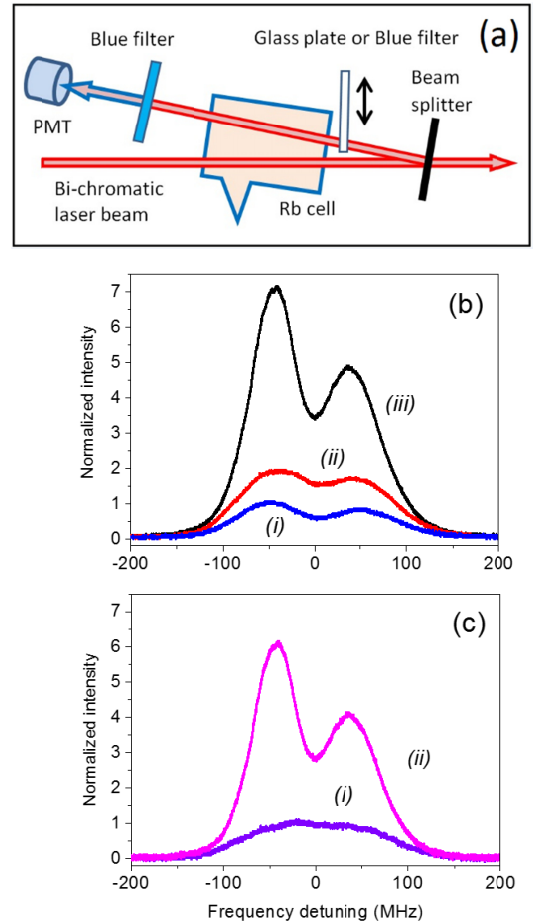

Fig. 3. (a) Optical scheme of the seeding experiment. (b) Experimentally measured blue light intensity as a function of 776 nm laser frequency detuning from the 5P$_{3/2}$(F'=4)-5D$_{5/2}$(F''=5) transition, while the fixed frequency of the 780 nm laser is tuned to the 5S$_{1/2}$(F=3)-5P$_{3/2}$(F'=4) transition. Curve (i) represents the intensity of first-passage CBL generated with blue filter inserted between the cell and the mirror. Curve (ii) shows the combined intensity profile of blue emission generated with inserted plain glass plate that blocks the mid-IR radiation only. Curve (iii) is recorded without any filter between the cell and the mirror. (c) Derived experimental signals proportional to the CBL intensity generated for the second passage only, (i) without and (ii) with IR seeding.

At the maximum atom number density N used in our experiment (N ≈ 2.0 ×10$^{12}$ cm$^{-3}$) the collisional broadening [20] is still small compared to the natural linewidth of all the optical transitions involved in the two-photon excitation and cascade decay.

Extended-cavity diode lasers (ECDL) at 780 and 776 nm are used for excitation of Rb atoms to the $5D_{5/2}$ level. Radiation from both ECDLs is combined on a non-polarizing beam splitter to form a two-colour beam. The polarizations and intensity of the 780 nm and 776 nm components are controlled with wave plates and polarizers. After transmission through an iris, the diameter of the parallel two-colour laser beam inside the Rb cell is ~1.2 mm, making the aspect ratio of the atom-light interaction region approximately 0.03. The maximum powers of the 780 and 776 nm components just before entering the cell are 8 and 5 mW, respectively.

Photodiodes and photomultiplier tubes (PMT) are used for blue and laser light detection. The 5.23 µm radiation is detected using a room-temperature photovoltaic detector based on a variable-gap HgCdTe semiconductor with 1 mm² light-sensitive area and a lock-in amplifier. For this the laser light at 780 nm is mechanically chopped at 400 Hz before entering the cell.

To obtain information about the spatial properties of the generated mid-IR radiation the 5.23 µm beam is scanned across the detector using an oscillating mirror or the detector is moved across the beam.

## 3. RESULTS AND DISCUSSION

### A. Seeding experiment

First, we consider the case when a blue glass filter is inserted between the Au-coated plate and the Rb cell. As the blue filter is highly opaque for both the IR and laser radiation and transparent only for blue light, the PMT signal represents the intensity of CBL generated in a single pass through the cell. For a qualitative understanding, we ignore the weak absorption of the CBL in the second passage. Curve (i) in Figure 3b demonstrates the blue light intensity as a function of frequency detuning of the 776 nm component of the two-colour laser. The two-peak CBL spectrum is familiar from previous studies [6, 11].

If the blue filter between the mirror and the cell is replaced with a plain glass plate, then the back-reflected beam contains attenuated laser light along with the first-passage CBL, while the mid-IR emission is blocked. The typical resonant absorption of the 780 nm and 776 nm components of the two-colour beam transmitted through the warm cell varies in the range of 80-90% and 20-25%, respectively. We find that despite such strong attenuation the reflected beam is still strong enough to generate new CBL during the second passage through the Rb vapour. In this case the PMT signal is proportional to the combined intensity of CBL produced in the double-pass arrangement. Its spectral dependence is depicted in Figure 3b by curve (ii). The intensity of the observed CBL is close to double that of the blue filter case despite the attenuation by the blue filter and the glass plate being similar.

Removing the plain glass plate between the mirror and the Rb cell has a much stronger effect on the CBL intensity. In this case the forward-directed blue light at 5.23 µm generated by the first passage enters the second interaction region along with the blue and laser light. The mid-IR radiation at 5.23 µm appears to act as a seed for emission on the population inverted the $5D_{5/2}$-$6P_{3/2}$ manifold produced in this region, which in turn results in a significant enhancement of the combined blue light intensity as curve (iii) reveals.

It is interesting to compare the maximum intensities and spectral dependences of CBL generated with and without seeding the second interaction region.

Curve (i) of Figure 3c is generated by subtracting the PMT signal recorded with the blue filter (curve (i) of Figure 3b) from the signal recorded with the plain glass plate (curve (ii) of Figure 3b) and so characterizes the intensity of blue light generated during the second passage only, with some blue light seeding but without injecting extra

IR radiation. Some CBL attenuation due to transmission through the blue filter is taken into account. The spectral profiles of CBL generated in the first and the second interaction regions without and with coherent blue light seeding (curves (i) in Figures 3b and 3c, respectively) are distinctly different. We attribute this to different power broadening produced by the resonant light at 780 nm.

The effect of seeding at 420 nm is of interest here, and the arrangement we use allows such seeding. We note that recently a ring cavity for blue light both enhanced the output power and narrowed the linewidth of the blue light generated by four wave mixing in a Rb vapour cell [21, 22]. Seeding the interaction region by coherent and directional blue light may establish a mutual coherence between the blue fields generated in the different passages, given that the optical frequencies of the CBL generated in different cells using the same pair of lasers have been observed to be different [6, 23]. However, our attempts to systematically seed with blue light have produced inconclusive results and further work is planned. Clearly, the enhancement and instabilities presented in Figure 2b are not due to seeding the interaction region by coherent and directional blue light.

The intensity profile of curve (ii) in Figure 3c is the difference between curve (iii) of Figure 3b, recorded with no plate between the cell and the mirror, and curve (i) of Figure 3b, recorded with the blue glass plate inserted. It thus represents the CBL generated in the second region with IR seeding. This approximately six-fold CBL intensity increase produced by injecting extra radiation at 5.23 µm emphasizes the importance of the process of amplified spontaneous emission that is responsible for frequency down-conversion, as the enhanced mid-IR light is directly involved in preparing nonlinear polarization for the $5S_{1/2}$ - $6P_{3/2}$ transition ($P \sim \chi^{(3)} E_1 E_2 E_{IR}$, where $\chi^{(3)}$ is nonlinear susceptibility and $E_i$ is the amplitude of the optical field at 780, 776 and 5 320 nm, respectively) and subsequent CBL generation. However, the mechanism of forward-directed mid-IR enhancement shown in Figure 2 remains to be clarified. It seems unlikely that the observed enhancement is due a low-finesse cavity effect produced by multiple reflections of mid-IR emission from both the cell windows because the enhancement is not critically dependent on the cell orientation. In the following, we suggest an alternative mechanism for the mid-IR radiation enhancement.

### B. Velocity selective and velocity insensitive two-photon excitations

To date, studies of the FWM process in diamond-type energy-level systems in alkali atoms, mostly in Rb, have focused almost exclusively on detecting the CBL and inferring details of the generation process from the blue light properties [1-13]. The primary reason for this is that the mid-IR radiation at 5.23 µm generated by the process of amplified spontaneous emission (ASE) is completely absorbed by the glass windows of vapour cells usually used in such experiments. However, the mid-IR step is a crucial component of the FWM process in this system and knowledge of its properties could provide complementary information that is invaluable for a proper understanding of the FWM process itself. Furthermore, the backward-directed mid-IR radiation is of particular interest because of possible applications in remote atmospheric sensing [23]. Up to now, few attempts have been made to examine the 5.23 µm emission. No evidence of backward-directed 5.23 µm emission was found in [2], though later such emission was detected and analysed [4, 16].

In the case of two lasers at 780 and 776 nm Rb atoms could be efficiently excited to the $5D_{5/2}$ level either by co- or counter-propagating beams. It was recently shown that over a wide range of experimental parameters the backward- and forward-directed ASE at 5.23 µm is stronger in the case of quasi Doppler-free, or velocity insensitive, two-photon excitation produced by counter-propagating beams [17]. It was also shown that the number of Rb atoms excited to the $5D_{5/2}$ level by two-colour resonant radiation in the co- and counter-propagating configurations could be approximately equal even with a ten times weaker counter-propagating beam at 776 nm, compared to the co-propagating beam at this wavelength. This suggests that efficient excitation to the $5D_{5/2}$ level, and subsequent ASE, might occur simply due to back-reflection of the applied laser fields from the cell windows, even without any externally applied counter-propagating light.

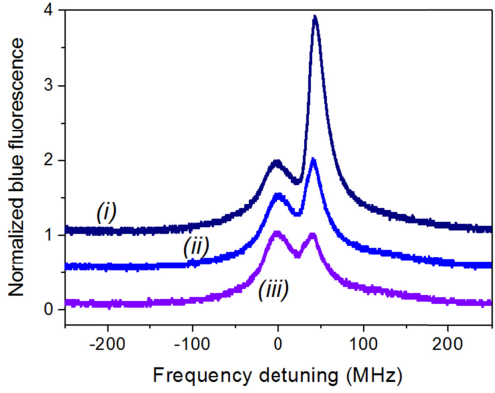

Fig. 4. Isotropic blue fluorescence at 420 nm as a function of frequency detuning of the 776 nm laser from the maximum velocity-selective two-photon excitation. Fluorescence profiles recorded at (i) 90%, (ii) 20% and (iii) 10% back-reflection. The fixed frequency 780 nm laser is red detuned from the $5S_{1/2}(F=3)$ - $5P_{3/2}(F'=4)$ transition, $\Delta\nu_{780} = \nu_{780} - \nu_{34} \approx$ -22 MHz.

At the two-photon resonance ($\nu_{780} + \nu_{776} = \nu_{FF''}$, where $\nu_{FF''}$ are the frequencies of the two-photon transition between the $5S_{1/2}(F=3)$ and $5D_{5/2}(F'')$ hyperfine levels) the counter-propagating laser beams can interact simultaneously with nearly the whole velocity distribution, as the Doppler shift is almost compensated when the wavelengths of the counter-propagating beams are very close. In the co-propagating configuration, the light-atom interaction occurs in a single velocity group. This could explain the observed higher ASE at 5.23 μm in the counter-propagating configuration. But, is the window-reflected component at 776 nm intense enough for efficient two-photon excitation?

To investigate this, we have made some measurements of isotropic blue fluorescence emitted by cascade-decaying Rb atoms for different experimental conditions, as the fluorescence intensity is a good indicator of the number of atoms excited to the $5D_{5/2}$ level.

We first demonstrate that counter-propagating beams could produce efficient excitation even despite a significant power imbalance. Figure 4 shows spectral profiles of isotropic blue fluorescence from Rb atoms excited by the two-colour laser beam transmitted though the cell and back-reflected from mirrors with different reflectivity in the near-IR spectral range.

Velocity-selective and velocity insensitive two-photon excitation produced by co- and counter-propagating laser light are easily distinguished in the spectral dependences if the 780 nm laser is frequency detuned from the $5S_{1/2}(F=3)$-$5P_{3/2}(F'=4)$ transition and $|\nu_{780} - \nu_{34}| < \Delta\nu_D$, where $\Delta\nu_D$ is the Doppler width [17].

If 90% of the laser light transmitted through the cell is reflected straight back, the maximum intensity of isotropic blue fluorescence is approximately three times higher compared to the velocity-selective peak at zero detuning of the 776 nm laser. Curve (iii) shows that the velocity selective and nearly Doppler-free two-photon peaks are almost equal at 10% back-reflection level. This confirms that much weaker counter-propagating laser beams at 776 nm together with the input 780 nm beam are able to excite a considerable number of Rb atoms to the $5D_{5/2}$ level.

## C. Cell tilt experiments

We now analyse the spectral dependences of isotropic blue fluorescence obtained at different tilts of the Rb cell with no counter-propagating light from external mirrors. In this case velocity insensitive two-photon excitation can originate only from the combined action of the 776 nm light back-reflected from the exit cell window and the 780 nm component of the applied two-colour laser beam. At large cell tilts ($\theta > 60$ mrad) the reflected beam from the cell exit window cannot play any role in the excitation process because of poor overlapping with the applied laser beam inside the cell. For this reason, Doppler-free two-photon excitation does not occur. As the overlapping increases at smaller cell tilts, the velocity insensitive two-photon peak of the blue fluorescence becomes noticeable at $\theta \approx 56$ mrad (curve (i) in Figure 5a). At $\theta \approx 6$ mrad and $\Delta\nu_{780} \approx$ -22 MHz it reaches the level of the velocity-selective peak.

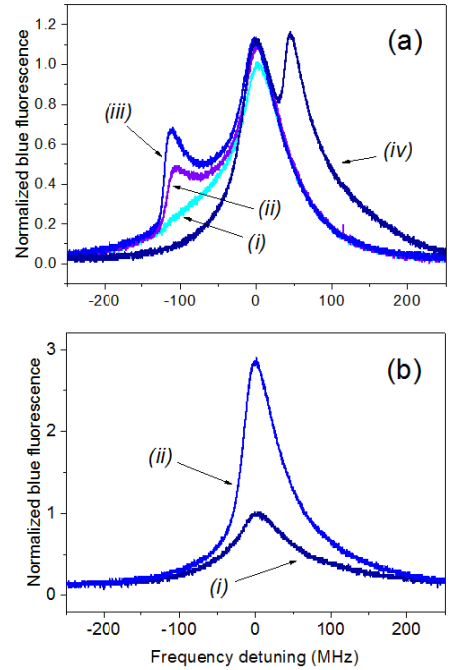

Fig. 5. Blue fluorescence as a function of frequency detuning of the 776 nm laser from the maximum velocity-selective two-photon excitation. (a) Fluorescence profiles recorded with no external mirror at different cell tilts: (i) $\theta = 56$ mrad, (ii) $\theta = 28$ mrad, (iii) $\theta = 14$ mrad and $\Delta\nu_{780} = \nu_{780} - \nu_{34} \approx$ +56 MHz; while curve (iv) is taken at $\theta = 6$ mrad and $\Delta\nu_{780} \approx$ - 22 MHz. (b) Fluorescence profiles taken at (i) $\theta = 56$ mrad and (ii) $\theta = 6$ mrad, while $\nu_{780} \approx \nu_{34}$.

The number of Rb atoms excited to the $5D_{5/2}$ level is higher when the fixed frequency laser component at 780 nm is precisely tuned to the cycling transition $5S_{1/2}(F=3)$-$5P_{3/2}(F'=4)$ and the cell tilt is small ($\theta \leq 8$ mrad). In this case the two fluorescence peaks merge as the resonant conditions for the stepwise and two-photon excitation are the same. Blue fluorescence taken at the large and small cell tilts, while the 780 nm laser is tuned to the $5S_{1/2}(F=3)$-$5P_{3/2}(F'=4)$ transition, are shown in Figure 5b. In the case of the small cell tilt, the regions of two-photon and stepwise excited atoms perfectly overlap which results in the at least 3-fold blue fluorescence enhancement.

Thus, the spectral dependences of the isotropic blue fluorescence presented in Figures 4 and 5 suggest that laser light reflected from the cell window could play a significant role in preparing population inversion on the $6P_{3/2}$-$5D_{5/2}$ transition through the mechanism of nearly Doppler-free two-photon excitation. This mechanism also dramatically modifies the conditions for ASE at 5.23 μm. We find that velocity-selective and nearly Doppler-free velocity-insensitive two-photon excitation could produce two spectrally and spatially distinguishable mid-IR fields, as the spectral and spatial resonant conditions for the processes are different.

Taking into account that the aspect ratio of the active region is approximately 0.03, we find that at intermediate cell tilts (10 mrad < θ < 40 mrad) the overlap is large enough for efficient excitation of Rb atoms to the $5D_{5/2}$ level by these counter-propagating laser fields. The line (i) in Figure 6a shows the direction of maximum overlap of the beams, which is therefore the direction of maximum gain on the $5D_{5/2}$-$6P_{3/2}$ transition. We note that this direction does not coincide with the direction of the maximum gain for the velocity-selective excitation case.

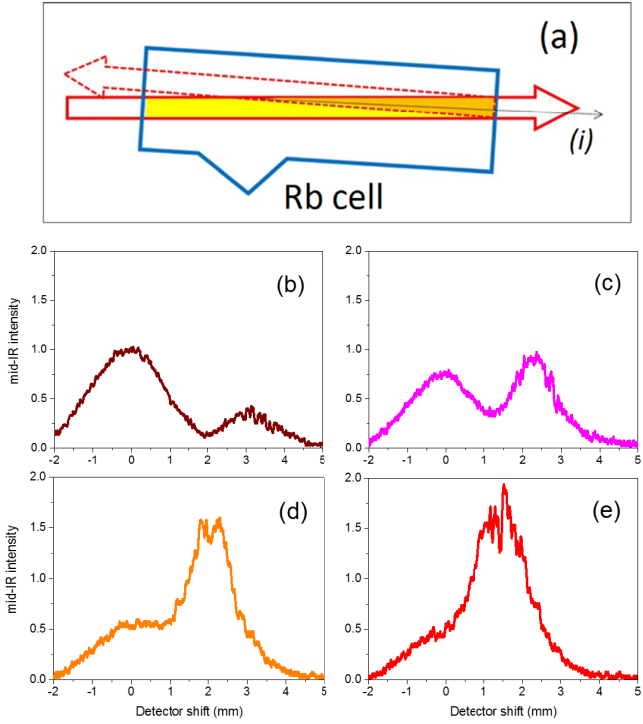

Fig. 6. (a) Sketch of two-colour laser beam propagation and reflection from the exit window of the tilted Rb cell. Yellow- and orange-marked areas indicate the regions containing the population-inverted Rb atoms excited by co- and counter-propagating laser fields, respectively. Black line (i) shows the direction of the maximum gain due to the nearly Doppler-free two-photon excitation. (b)-(e) Spatial profiles of mid-IR profiles of forward-directed ASE at 5.23 μm at different Rb cell tilts (θ ≈ 35; 30; 25 and 18 mrad respectively) obtained by shifting the detector across the IR beam approximately 9 cm from the exit cell window. The 780 nm and 776 nm lasers are tuned to the $5S_{1/2}(F=3)$-$5P_{3/2}(F'=4)$ and $5P_{3/2}(F'=4)$-$5D_{5/2}$ transitions, respectively. The plotted mid-IR profiles are normalized relative to the level of the 5.23 μm emission at the large cell tilt (θ ≈ 60 mrad).

The initially isotropic spontaneous emission for the $5D_{5/2}$-$6P_{3/2}$ transition is predominantly amplified along the direction of maximum column density of population-inverted atoms. The yellow- and orange-marked sections in Figure 6a represent the regions that contain Rb atoms excited to the $5D_{5/2}$ level by the co-propagating and counter-propagating laser fields, respectively.

At very small cell tilts (θ < 10 mrad) the forward-directed mid-IR emission is also co-propagating with the applied laser beam. In this case the two regions inside the cell where the population inversion is produced either by the velocity-sensitive stepwise or velocity-insensitive two-photon processes merge.

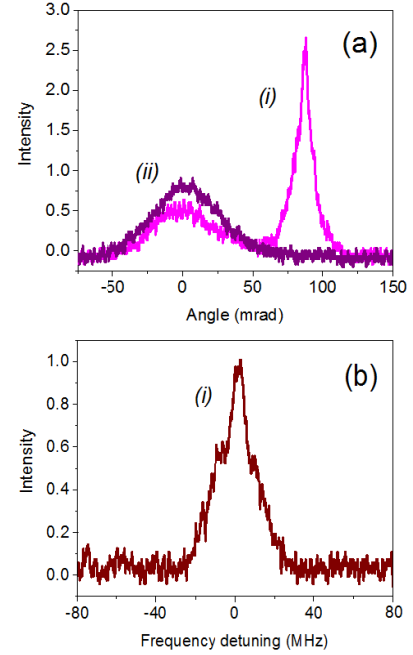

Fig. 7. (a) Spatial profiles of forward-directed IR emission taken by scanning the mid-IR beam across the detector. Curve (i) demonstrates the co-existence of two IR fields originating from velocity-insensitive and velocity-selective two-photon excitation (narrower and wider peaks, respectively) recorded in the tilted Rb cell (θ = 40 mrad) and at exact frequency tuning ($\nu_{780} = \nu_{34}$ and $\nu_{780} + \nu_{776} = \nu_{35}$). If both lasers are detuned from the corresponding transitions ($\nu_{780} ≈ \nu_{34}$ - 45 MHz) ASE occurs only in the co-propagating direction as curve (ii) shows. (b) Forward-directed mid-IR intensity as a function of frequency detuning of the 776 nm laser. Curve (i) demonstrates IR profile recorded for the small cell tilt (θ = 6.4 mrad) and exact frequency tuning ($\nu_{780} = \nu_{34}$ and $\nu_{780} + \nu_{776} = \nu_{35}$).

At some angles θ the ASE process in the co-propagating direction is partially suppressed because of competition with the off-axis ASE process. Experimental curves shown in Figure 6(b-e) demonstrate evolution of the forward-directed mid-IR spatial profiles with the Rb cell tilt reduction. The profiles are taken with both lasers are tuned precisely and actively locked to the corresponding $5S_{1/2}(F=3)$ - $5P_{3/2}(F'=4)$ and $5P_{3/2}(F'=4)$ - $5D_{5/2}(F''=5)$ transitions. The off-axis component originating from the velocity-insensitive two-photon process dominates at the cell tilts 15 mrad < θ < 30 mrad. The spatial distribution of the off-axis component is narrower. We explain this as a result of a more homogeneous longitudinal distribution of the population-inverted Rb atoms produced by the nearly Doppler-free two-photon excitation in the pencil-shaped atom-light interaction region. The higher intensity noise of the off-axis mid-IR emission could be attributed to an effect of interference between multiple reflections from the cell windows. This issue as well as mutual coherence of the on- and off-axis components require further investigation.

We note that the co-existence of the on- and off-axis mid-IR beams and their relative intensities depend on the frequency detuning of the applied two-component laser radiation from the corresponding one-photon transitions. Figure 7a shows the spatial profiles of forward-directed IR emission at the two-photon resonance ($v_{780} + v_{776} = v_{FF}$", where $v_{FF}$" are the frequencies of the two-photon transition between the $5S_{1/2}$(F=3) and $5D_{5/2}$(F") hyperfine levels, respectively, while the fixed frequency component at 780 nm is tuned precisely to the $5S_{1/2}$(F=3) -$5P_{3/2}$(F'=4) transition ($v_{780} = v_{34}$), or red-detuned from the resonance ($\Delta v_{780} = v_{780} - v_{34} \approx$ - 45 MHz). The on-axis peaks originating from the velocity-selective excitation produced by the co-propagating components are wider and present in both cases, whereas the more collimated off-axis emission due to the nearly Doppler-free two-photon excitation occurs only at the exact resonances of both the laser fields. In this case the stepwise and two-photon excitation is indistinguishable.

The spectral dependence of the forward-directed mid-IR emission at small cell tilt is shown in Figure 7b, where the horizontal axis is detuning of the 776 nm laser from the $5P_{3/2}$(F'=4) - $5D_{5/2}$ transition and the 780 nm laser is ~~either~~ tuned to the $5S_{1/2}$(F=3) -$5P_{3/2}$(F'=4) transition ($v_{780} = v_{34}$). The shape of the IR resonance suggests that the detected signal is the sum of the independent fields generated by population-inverted atoms within either a certain velocity group or from the whole velocity distribution. A detailed study of spatial properties of the backward and forward-directed ASE is the subject of a following paper.

## 4. CONCLUSION

We have shown how the orientation of a Rb cell can significantly affect the intensity and spectral characteristics of both the frequency up- and down-converted fields generated as a result of nonlinear processes in Rb vapour.

By seeding the interaction region with polychromatic resonant light, we have revealed the important role of the process of amplified spontaneous emission that is responsible for generation of directional radiation at 5.23 μm.

We have shown that the process of velocity-insensitive Doppler-free two-photon excitation is central for understanding the observed CBL and mid-IR light enhancement.

The velocity-selective and velocity-insensitive two-photon excitation for some experimental conditions can produce two spectrally and spatially distinguishable mid-IR fields at 5.23 μm.

The presented results give insight into the origin of the wave mixing process in alkali vapours and should be useful for explaining the distinctive spectral dependences of backward- and forward-directed ASE at 5.23 μm reported in [4].

**Funding Information.** This work is partially supported by the US Office of Naval Research, Global (N62909-16-1-2113).

**Acknowledgment**. We thank Peter Hannaford for useful discussions.

## References

1. A. S. Zibrov, M. D. Lukin, L. Hollberg, and M. O. Scully, "Efficient frequency up-conversion in resonant coherent media", Phys. Rev. A **65**, 051801 (2002).
2. A. S. Zibrov, L. Hollberg, V. L. Velichansky, M. O. Scully, M. D. Lukin, H. G. Robinson, A. B. Matsko, A. V. Taichenachev, and V. I. Yudin, "Destruction of Darkness: Optical Coherence Effects and Multy-Wave Mixing in Rubidium Vapor," Proceedings of the 17-th International Conference on Atomic Physics, E. Arimondo, P. DeNatale, and M. Inguscio, eds. (AIP Conference Proceedings **551**, 204 (2001).
3. J. F. Sell, M. A. Gearba, B. D. DePaola, and R. J. Knize, "Collimated blue and infrared beams generated by two-photon excitation in Rb vapor" Opt. Lett. **39**, 528 (2014).
4. A. Akulshin, D. Budker, and R. McLean, "Directional infrared emission resulting from cascade population inversion and four-wave mixing in Rb vapor" Opt. Lett. **39**, 845 (2014).
5. T. Meijer, J. D. White, B. Smeets, M. Jeppesen, and R. E. Scholten, "Blue five-level frequency-upconversion system in rubidium" Opt. Lett. **31**, 1002 (2006).
6. A. M. Akulshin, R. J. McLean, A. I. Sidorov, and P. Hannaford, "Coherent and collimated blue light generated by four-wave mixing in Rb vapour", Optics Express **17**, 22861 (2009).
7. A. Vernier, S. Franke-Arnold, E. Riis, and A. S. Arnold, "Enhanced frequency up-conversion in Rb vapour", Optics Express **18**, 17020 (2010).
8. J. T. Schultz, S. Abend, D. Döring, J. E. Debs, P. A. Altin, J. D. White, N. P. Robins, and J. D. Close, "Coherent 455 nm beam production in a cesium vapour", Opt. Lett. **34**, 2321 (2009).
9. G. Walker, A. S. Arnold, and S. Franke-Arnold, "Trans-spectral orbital angular momentum transfer via four-wave mixing in Rb vapour", Phys. Rev. Lett. **108**, 243601 (2012).
10. C. V. Sulham, G. A. Pitz, and G. P. Perram, Appl. Phys. B **101**, 57 (2010).
11. E. Brekke and L. Alderson, "Parametric four-wave mixing using a single cw laser", Opt. Lett. **38**, 2147 (2013).
12. A. Akulshin, C. Perrella, G.-W. Truong, A. Luiten, D. Budker, and R. McLean, Appl. Phys. B **117**, 203 (2014).
13. E. Brekke and E. Herman, "Frequency characteristics of far-detuned parametric four-wave mixing in Rb", Opt. Lett. **40**, 5674 (2015).
14. A. M. Akulshin, B. V. Hall, V. Ivannikov, A. A. Orel, and A. I. Sidorov, "Doppler-free twophoton resonances for atom detection and sum frequency stabilization" J. Phys. B: At. Mol. Opt. Phys. **44**, 215401 (2011).
15. A. G. Radnaev, Y. O. Dudin, R. Zhao, H. H. Jen, S. D. Jenkins, A. Kuzmich, T. A. B. Kennedy, "A quantum memory with telecom-wavelength conversion", Nature Phys. **6**, 894 (2010).
16. A. Akulshin, D. Budker, B. Patton, and R. McLean, "Nonlinear processes responsible for mid-infrared and blue light generation in alkali vapours," arXiv:1310.2694 [physics. atom-ph] (2013).
17. A. M. Akulshin, N. Rahaman, S. A. Suslov, and R.J. McLean, "Amplified spontaneous emission at 5.23 μm in two-photon excited Rb vapour", arXiv: 1701.00232 [physics. atom-ph] (2017).
18. http://mellesgriot.com/Frontend/PDFs/TechGuide.pdf.
19. Rochester Scientific; http://rochesterscientific.com/ADM.
20. A. M. Akulshin A. A. Celikov, V. A. Sautenkov, T. A. Vartanian and V. L. Velichansky, "Intensity and concentration dependence of Doppler-free resonance in selective reflection", Opt. Comm. **85**, 21 (1991).
21. R. F. Offer, J. W. C. Conway, E. Riis, S. Franke-Arnold, and A. S. Arnold, "Cavity-enhanced frequency up-conversion in rubidium vapour", Opt. Lett. **41**, 2177 (2016).
22. E. Brekke and S. Potier, "Optical cavity for enhanced parametric four-wave mixing in rubidium", Appl. Opt. **56**, 46 (2017).
23. A. Akulshin, C. Perrella, G.-W. Truong, R. McLean, and A. Luiten, "Frequency evaluation of collimated blue light generated by wave mixing in Rb vapour", J. Phys. B: At. Mol. Opt. Phys. 45, 245503 (2012).
24. J. Kasparian and J.-P. Wolf, "Physics and applications of atmospheric nonlinear optics and filamentation", Optics Express, **16**, 466 (2008).